\documentclass{PoS}
\usepackage{graphicx}
\usepackage{amsmath}
\usepackage{amsfonts}
\usepackage{graphics}
\usepackage[authoryear,round]{natbib}
\newcommand\aj{AJ}%
\newcommand\araa{ARA\&A}%
\newcommand\apj{ApJ}%
\newcommand\apjl{ApJ}%
\newcommand\apjs{ApJS}%
\newcommand\apss{Ap\&SS}%
\newcommand\aap{A\&A}%
\newcommand\mnras{MNRAS}%
\newcommand\pasa{PASA}%
\newcommand\rmxaa{Rev. Mexicana Astron. Astrofis.}%
\newcommand\nat{Nature}%
\newcommand\iaus{IAU~Symp.}%
\newcommand\farcs{\mbox{$.\!\!^{\prime\prime}$}}%
\newcommand\ion[2]{#1$\;${\small\rmfamily\@Roman{#2}}\relax}%
\newcommand\sun{\odot}%
\title{SKA studies of nearby galaxies: star-formation, accretion processes and molecular gas across all environments}

\ShortTitle{Nearby galaxies with the SKA}

\author{\speaker{R. J. Beswick}$^1$\thanks{Robert.Beswick@manchester.ac.uk},
  E. Brinks$^2$, M. A. P\'erez-Torres$^3$,
   A. M. S. Richards$^1$, S. Aalto$^4$, A. Alberdi$^2$,
	M. K. Argo$^1$, I. van Bemmel$^{5}$, J. E. Conway$^4$, C. Dickinson$^1$
	D. M. Fenech$^6$, M. D. Gray$^1$, 
	H-R Kl\"ockner$^7$, E. J. Murphy$^8$,
	T. W. B. Muxlow$^1$, M. Peel$^1$, A. P. Rushton$^{9,10}$, 
	E. Schinnerer$^{11}$\\
$^1$Jodrell Bank Centre for Astrophysics/{\it e-}MERLIN, The
        University of Manchester, M13~9PL, UK\\
$^2$Centre for Astrophysics Research, University of Hertfordshire,
	AL10~9AB, UK\\
$^3$Instituto de Astrofisica de Andalucia (IAA-CSIC), E-18008
	Granada, Spain\\
$^4$Department of Earth and Space Sciences, Chalmers University of
        Technology, Onsala Space Observatory, SE-439 92, Onsala, Sweden\\
$^{5}$Netherlands Institute for Radio Astronomy (ASTRON), Postbus-2,
        NL-7990 AA Dwingeloo, The Netherlands 

$^6$Department of Physics and Astronomy, University College London,
	London, WC1E~6BT, UK\\
$^7$Max-Planck-Institut f\"ur Radioastronomie, Auf dem H\"ugel 69,
	D-53121 Bonn, Germany\\
$^8$IPAC, Caltech, MC 220-6, Pasadena CA, 91125, USA \\
$^9$Department of Physics, Astrophysics, University of Oxford, Keble
        Road, Oxford, OX1~3RH, UK \\
$^{10}$School of Astronomy and Physics, Univeristy of Southampton,
        highfield, Southampton, SO17~1BJ, UK\\
$^{11}$Max-Planck-Institut f\"ur Astronomie, K\"onigstuhl 17, D-69117,
	Heidelberg, Germany\\
}

\abstract{The SKA will be a transformational instrument in the study
  of our local Universe. In particular, by virtue of its high sensitivity
  (both to point sources and diffuse low surface brightness emission), angular
  resolution and the frequency ranges covered, the SKA will undertake a
  very wide range of astrophysical research in the field of nearby
  galaxies. By surveying vast
  numbers of nearby galaxies of all types with $\mu$Jy sensitivity and
  sub-arcsecond angular resolutions at radio wavelengths, the SKA will
  provide the cornerstone of our understanding of star-formation and
  accretion activity in the local Universe. In this chapter we outline
  the key continuum and molecular line science areas where the
  SKA, both during phase-1 and when it becomes the full SKA, will have
  a significant scientific impact.

}

\FullConference{
Advancing Astrophysics with the Square Kilometre Array\\
June 8-13, 2014\\
Giardini Naxos, Italy}

\begin{document}

\section{Introduction}
The appearance of galaxies, and by this virtue, our Universe as a
whole is dominated by two physical processes: star-formation  and
accretion. Star-formation (SF) is fundamental to the formation and
evolution of galaxies, whilst accretion provides a major power source
in the Universe, dominating the emission from distant quasars down to
nearby X-ray binary systems. The feedback between these two processes
is also crucial, e.g., in reconciling the observed galaxy luminosity
function with predictions from the standard hierarchical clustering
models. Radio observations provide by far the best single diagnostic
of these two processes, allowing a direct view of SF even in dusty
environments and the detection of AGN and measurement of their
accretion rate at bolometric luminosities far below anything
detectable at higher energies. 

In this chapter we outline the scientific motivation for undertaking
large surveys and deep observations of
galaxies in the local Universe across all available frequency bands of
SKA1 and look forward to the full SKA. The high sensitivity, resolution and
imaging fidelity capabilities of SKA1, and the expected
enhanced capabilities of the full SKA,
will ensure that the SKA will become a dominant instrument for the detailed study of nearby galaxies
over the coming decades. 

Continuum surveys, undertaken during phase-1 covering the expected SKA dish
(MID/SUR) frequency ranges of 0.35--3 GHz and higher \citep[see
  also][in this proceedings]{murphy2014}, will provide $\mu$Jy sensitivities with exquisite image fidelity
over a wide range of spatial scales for all nearby galaxies. This will produce a complete census of SF and AGN activity as a function of galaxy mass, morphology and spectral type, black-hole mass
and luminosity; alongside optical/IR/mm surveys and observations from
ALMA, LSST, VISTA,
{\it Spitzer}, {\it Herschel} and {\it WISE}, this will be the cornerstone of multi-wavelength studies of the local Universe.

By providing high angular resolution observations ($\sim$0\farcs5 or better) the SKA will be able to decompose individual galaxies
into their compact radio source populations comprising accretion dominated AGN plus the tracers of
the early stages of SF, such as compact H{\sc ii} regions, super star clusters (SSCs), as well as
stellar end-points like X-ray binaries, planetary nebulae, supernovae (SNe) and their remnants (SNR).
In nearby galaxies the SKA will image these sources at physically important size scales
($\sim$few to tens of parsecs) characterising
individual sources and providing a detailed extinction-free census of the compact SF products
across a wide range of galaxy types and the environment parameter space they inhabit. For example, at
a sensitivity of a few $\mu$Jy at $\sim$GHz frequencies, a census of this type will detect all of the long-lived radio
SNR within several tens of Mpc, thus providing a measure of SF rates (SFRs) within local galaxies,
independent of the IR-radio correlation and obscuration
corrections. The mismatch between the measured Core Collapse SN
(CCSNe) rates and the cosmic massive SFR
\citep{mattila12} should be corrected. Such a measure will preferentially
trace high mass SF (M$>$8M$_\odot$), providing a direct tracer of the upper part of the
galaxy Initial Mass Function (IMF). The SKA will detect all AGN activity in the local Universe. By resolving and measuring the AGN and SF contributions, local Universe studies
of galaxies with the SKA will be vital in determining the contributions and interplay of these two key
physical processes which are critical for our interpretation of the high redshift Universe. On larger scales the SKA
will image the diffuse radio emission structure of nearby galaxies across a range of frequencies (few
hundred MHz -- several GHz, ideally up to at least 13GHz) thus allowing the separation of synchrotron and
thermal radio emission. The component separation of the thermal
emission of galaxies provides another unobscured measure of the massive SF activity
spanning a diverse range of galaxy types and environments, and providing a local Universe reference
point.

Observations of this type will be undertaken both commensally with other wide-area or
pointed observations \citep{prandoni2014}, along with specific targeted pointings focusing on this science theme.
Complementing the continuum science for nearby galaxies, high
sensitivity cold neutral, molecular and ionised ISM observations (the fuel for the SF and accretion processes traced via
continuum observations, see section\,{\ref{gas},  also  see
  \citealt{deblok2014, paladino2014,oonk2014} - this proceedings), and polarization
  observations of galaxies \citep[][also in these proceedings]{beck2014, heald2014}
will also be made by the SKA. This combined information will allow the
SKA to fully characterise the ongoing physics and fueling of SF and
accretion within the nearby
galaxy population.

\section{Radio continuum tracers of star-formation in the local Universe}

The SF history, along with current levels of SF, within
individual galaxies still remains a crucial physical parameter that observations are only now beginning to
accurately characterise. Traditionally, observations of optical line, IR and the global synchrotron emission
from galaxies have been used as proxies for SF. However, these tracers have some fundamental
flaws. In many galaxies optical emission lines are heavily obscured towards their centres, this can require
potentially large corrections, whilst IR emission, which essentially traces the light from young stars
reprocessed by dust and re-radiated at longer wavelengths, relies upon empirical interpretation of physically
complex processes in order for it to be related to SF. Global radio synchrotron emission provides an
alternative, and extinction-free, indication of SF, however, the link between radio synchrotron emission
and SF is also via complex physical mechanisms (see
section\,\ref{localvol}) which are generally calibrated using the
radio-to-IR correlation \citep{yun01,bell03,beswick08}. The long-standing critical issue
which remains is how to calibrate either global radio or IR emission as a measure of SF, and how this
calibration varies as a function of galaxy properties (e.g. gas
content, SFR, specific SFR, interaction/merger state) and their
global environment \citep{ken98}.

\begin{figure}
  \begin{tabular}{l}
    \includegraphics[width=0.33\linewidth,clip]{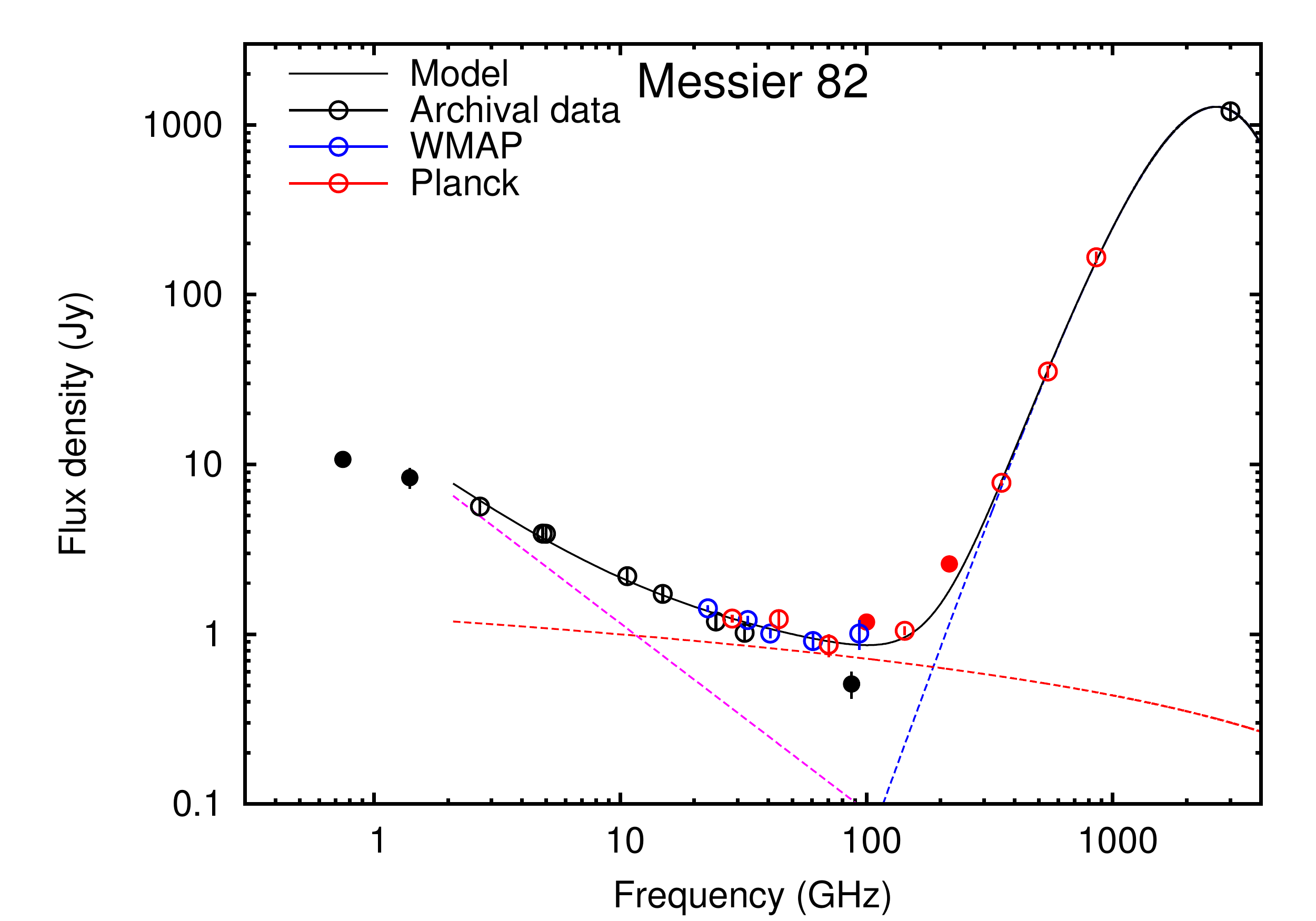}
 \includegraphics[width=0.33\linewidth,clip]{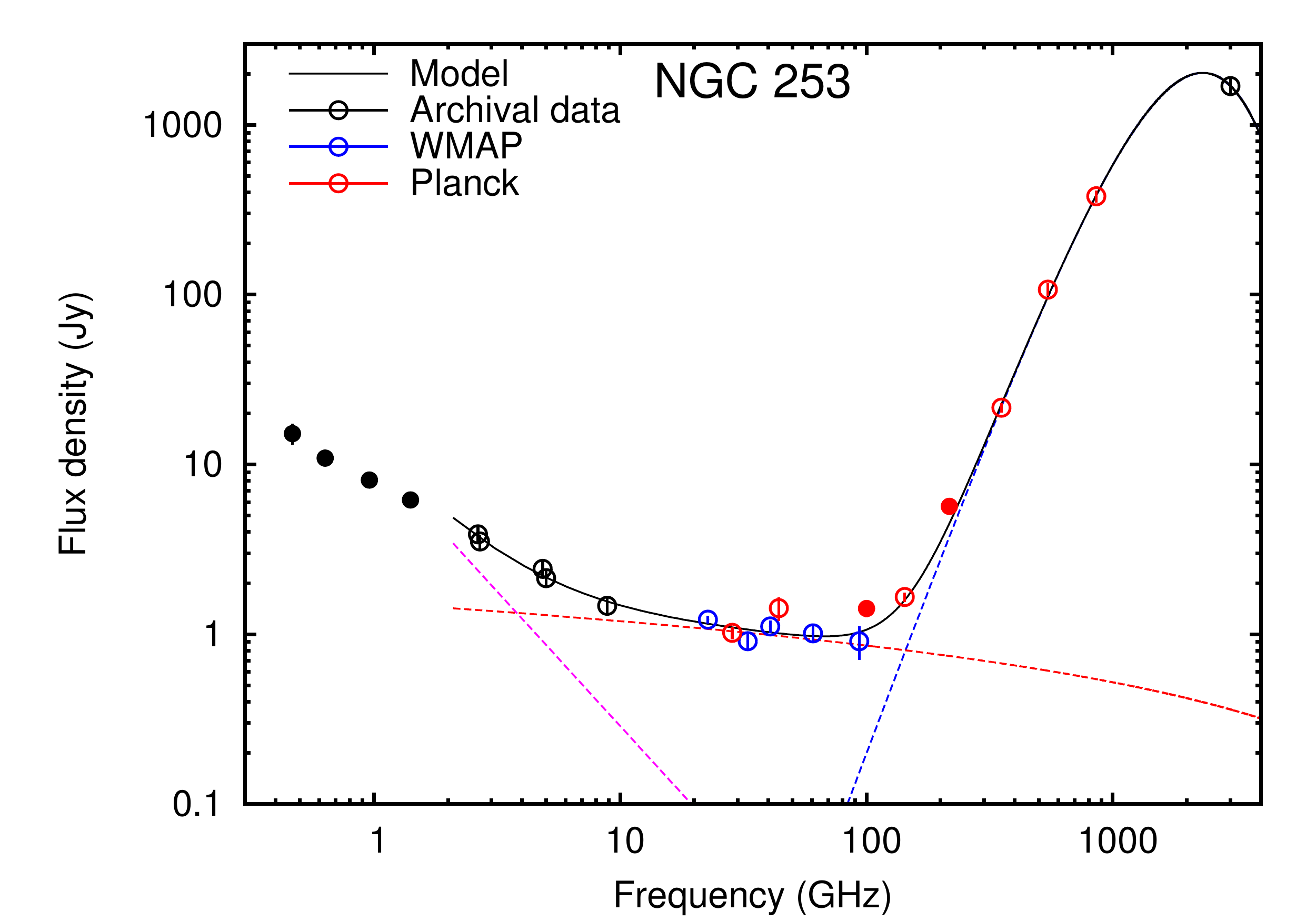}
 \includegraphics[width=0.33\linewidth,clip]{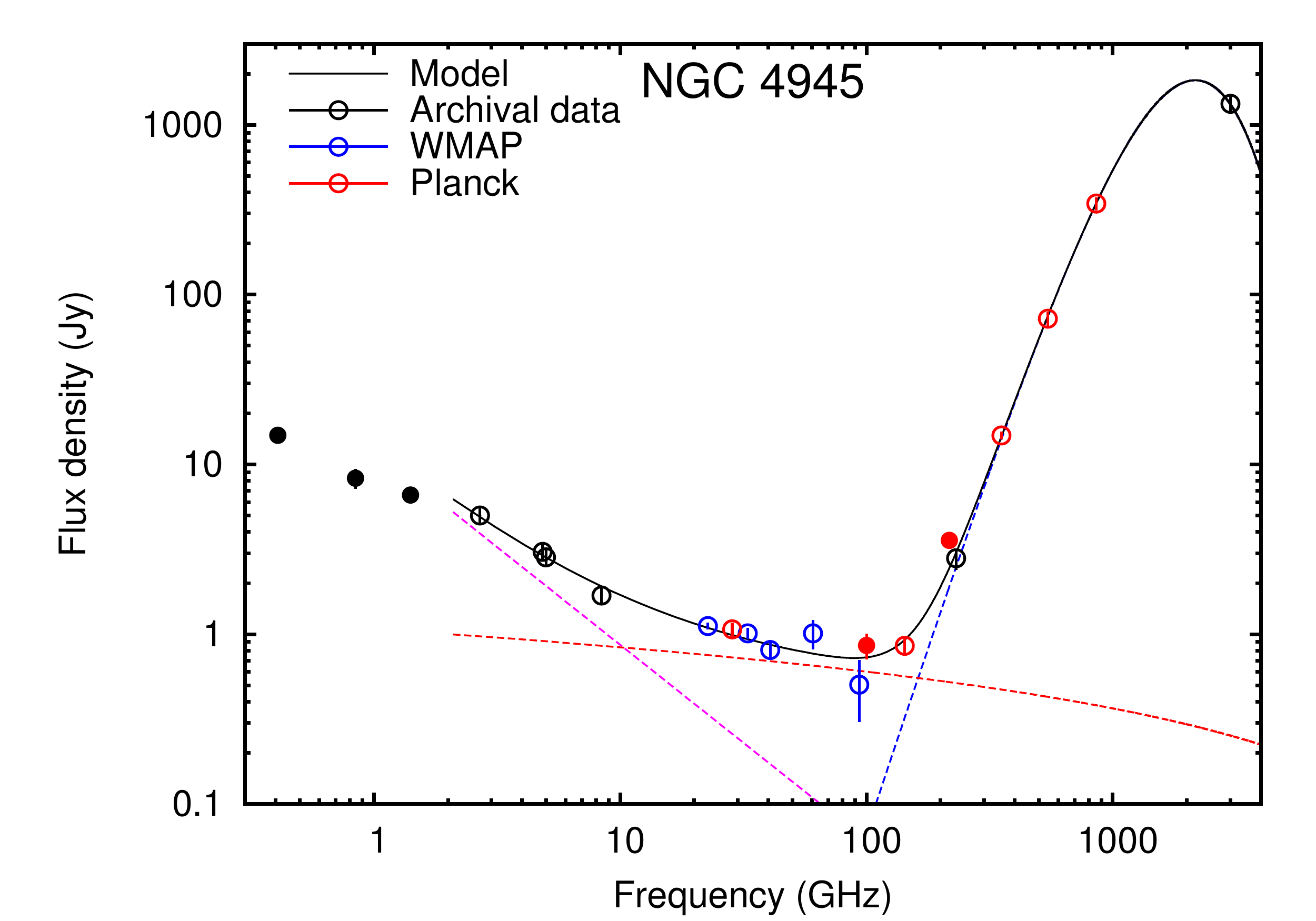}

  \end{tabular}
  \caption{\footnotesize{Spectral energy distribution of the nearby
      star-forming galaxies M82, NGC253 and NGC4945. Spectral fits to
      synchrotron (magenta), thermal (free-free - red) and thermal
      (dust - blue) are shown alongside the combination of these
      products as a solid line \citep{peel11}.}}
  \label{FigureSED}
\end{figure}

\subsection{Thermal and non-thermal radio emission as star-formation tracers}

\subsubsection{Radio-Infrared correlation and star-formation}
\label{localvol}
The physical underpinning of the radio--continuum (RC) ---
Far--Infrared (FIR) correlation is broadly understood to be the
formation of massive stars, which can believably be associated with
the dominant contributions to both the FIR and RC emission of
galaxies. The IR contribution is due to the hot dust envelopes
surrounding the massive stars, which are energised by the copious
amounts of Lyman continuum radiation that those stars emit. The radio
contribution includes two components (see Fig.\,\ref{FigureSED}). 
The first is the thermal continuum, free-free emission, which predominantly comes from H{\sc ii} regions, associated with massive  (M $\geq 8$ M$_{\odot}$) stars.
The second is the non-thermal, synchrotron emission from relativistic
electrons.  These electrons are accelerated  in the SNe
shocks that occur when massive stars explode as CCSNe (type Ibc and
type II SNe).  When the supernova blast wave interacts with the outer
material a shell is formed, and radio emission is generated initially
due to the interaction with the circumstellar medium (CSM), fading
away with time as the supernova continues to expand. Eventually, the
supernova shock meets the uniform interstellar medium (ISM), at which
time the SN suffers a re-brightening and continues emitting at radio
wavelengths for thousands of years, now as a SNR, which is also a very efficient cosmic ray electron (CRe) accelerator.  If all CRe lose their energy exclusively within the galaxy, it can be considered a calorimeter \citep{voelk89,lis96}.

Both continuum thermal free-free radio emission and synchrotron
radio emission can be considered as SF tracers. Thus if the presence of an AGN can be excluded (or its contribution neglected), one could then obtain independent estimates of the SF rate, and thus check the linearity of the RC(thermal)--FIR and RC(sync)--FIR correlations.

\begin{figure}
  \begin{tabular}{c}
    \includegraphics[width=\linewidth,clip]{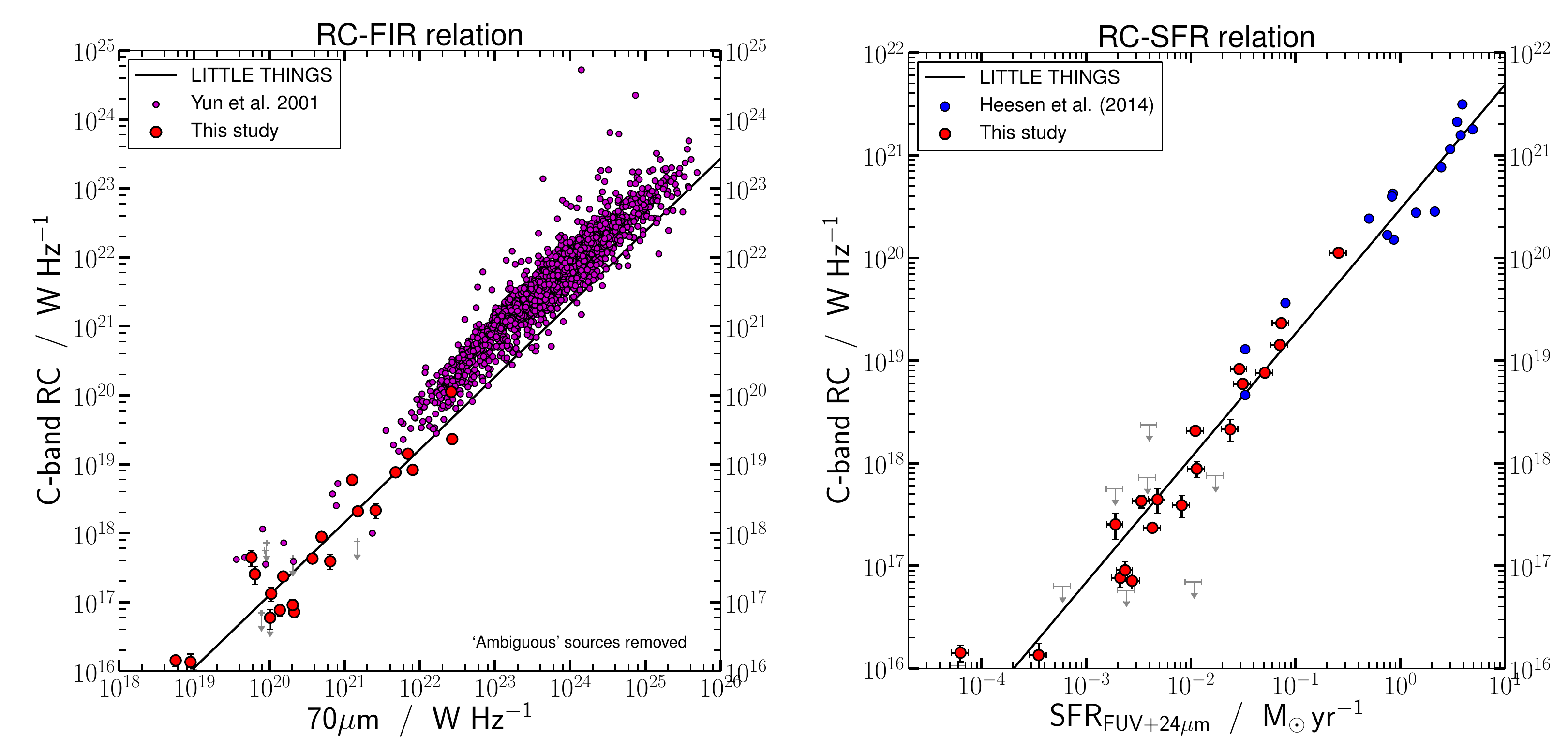}
  \end{tabular}
  \caption{\footnotesize{{\em left: RC--FIR relation} Filled purple
      symbols are integrated values for the galaxies taken from the
      sample collected by \citet{yun01}. Their VLA 1.4\,GHz data have
      been corrected to what would be expected at 6\,GHz and their
      IRAS $60\,\mu$m data to the Spitzer $70\,\mu$m passband. The red
      circles and upper limits (downward arrows) come from the 6\,cm
      VLA survey of 40 dwarf irregular galaxies by \citet{kit14} and
      are listed as ``This study" in the legend. The line is a fit
      through the red circles and corresponds to a power law slope of
      $0.9\pm0.1$ and a dispersion of 0.3\,dex. {\em right: RC--SFR
        relation} The blue circles are from \citet{hees14} who mainly
      studied spiral galaxies observed with the WSRT at 22\,cm from
      the SINGS survey. The WSRT 22\,cm fluxes have been corrected to
      6\,GHz. The red circles and upper limits (downward arrows) are
      from \citet{kit14}.}}
  \label{figure:RC-relation}
\end{figure}

RC and FIR emission both
depend on (recent) SF and will thus be correlated. Surprisingly, this
RC--FIR relation of galaxies holds over 4 orders of magnitude in
luminosity, irrespective of galaxy type
\citep{hel85,dejong85,beswick08}; displays only a 0.26\,dex scatter,
and has been observed to hold out to a redshift of about 3
\citep{garrett02,app04}. Fig.~\ref{figure:RC-relation} (left hand panel) taken
from \citet{kit14}, which is based on data from \citet{yun01}, shows the
RC--FIR relation based on the integrated emission from star forming
(spiral) galaxies. The RC--FIR relation can be understood if not only
the CRe lose all their energy within the galaxy but also the dust
absorbs and transforms all the emission related to recent SF into
FIR. More realistically, a galaxy is a leaky box, and models become
far more involved \cite[e.g.,][]{bell03,beswick08,lac10}. Despite the leaky
nature of galaxies, the relative successes of theoretical models lend
confidence that massive SF and RC emission are indeed
closely tied together. A well calibrated RC--FIR relation offers 
a powerful method to probe the cosmic SFR out to intermediate
redshifts, initially with SKA pathfinders and precursors, and
eventually with the SKA
\cite[e.g.,][]{mur09,lac10,mur12,norris13}. This will be an essential
prerequisite for deep extragalactic surveys
with the SKA since measuring the cosmic SF history will be
limited by our understanding of the radio continuum-to-SF
relationship rather than the number count statistics which limit today's observations.

Rather than relying on the FIR, which at best
provides modest angular resolution and relies on the availability of
suitable satellites, one can use the RC to determine directly the
current SFR in galaxies \cite[see][for a review]{con92}. The thermal
and non--thermal  emission are the result of fundamentally different
processes, with different RC--SFR relations expected for each
component. The thermal RC is expected to be directly proportional to
the SFR, depending as it does on the ionised flux from massive
stars. This makes it an ideal, virtually extinction--free proxy for SF
\citep{mur11}. The non--thermal RC depends on the magnetic field
strength as well as the cosmic--ray energy density, unless one assumes
an electron calorimeter which is unlikely, particularly for dwarf
galaxies, or in general those galaxies with large--scale
outflows. Usually one assumes energy equipartition between the CRe and
the magnetic field, so that the RC--SFR relation is closely connected
to a relation between the magnetic field and gas. \citet{pri92}
separated the RC into its thermal and non--thermal components and
found both to follow their own unique RC--FIR relation; the thermal
relation being indistinguishable from a one--to--one relation (power
law slope of $0.97\pm0.02$) whereas the non--thermal relation is
steeper, at $1.33\pm0.10$.  Although it is less obvious why the non--thermal RC would form a tight
relation with SF, empirically it does and  recent theoretical work
suggests the non--thermal RC--SFR follows a super--linear
\cite[e.g.,][]{nik97,schlei13} relation with a slope of
RC$\sim$SFR$^{1.3}$. Given that, at the lower frequencies, non--thermal
emission related to SF can be 1--2 orders of magnitude brighter than
the thermal counterpart, there is great potential in arriving at a
well calibrated and understood RC-SFR relation. A recent compilation
is shown in Fig.~\ref{figure:RC-relation} (right hand panel) taken
from \citet{kit14} which is based on data from \citet{hees14}. It shows
the RC--SFR relation based on the integrated emission from star
forming galaxies. 

\begin{figure}
  \begin{tabular}{c}
    \includegraphics[width=\linewidth,clip]{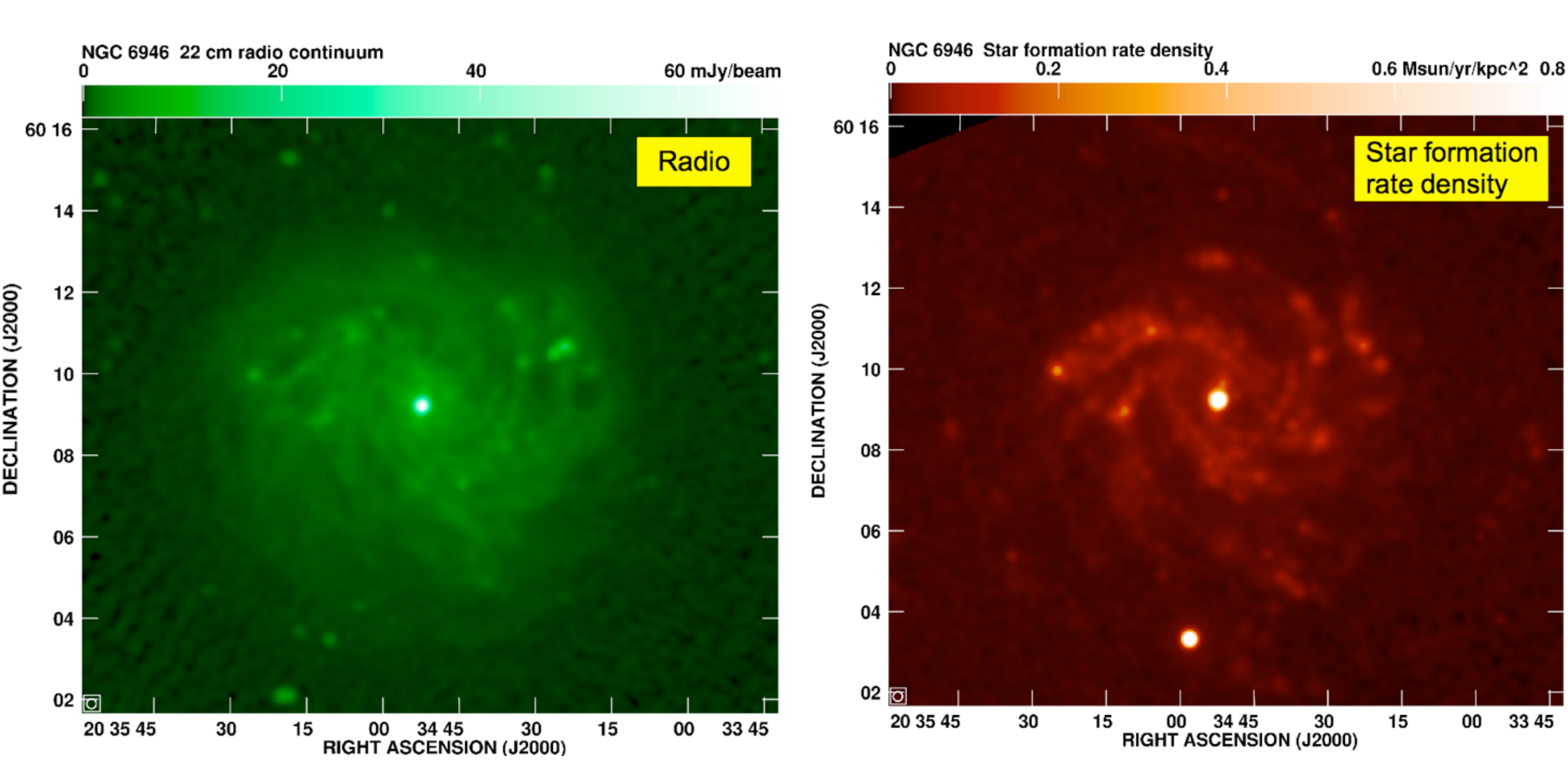}
  \end{tabular}
  \caption{\footnotesize{Images of the 22\,cm radio continuum and
      derived star-formation rate density for the nearby face-on
      star-forming galaxy NGC\,6946.}}
  \label{figure:RC-resolved}
\end{figure}

\subsubsection{From existing capabilities to the SKA}
Upgrades to existing instruments (in the form of wide--band end-to-end
systems)
have opened up a number possibilities including i) spatially resolved imaging of
dozens of spirals in the Local Universe down to $\mu$Jy
levels, ii) extending those studies to the low mass, low metallicity
regime of dwarf irregular galaxies, and iii) mapping of nearby galaxies at
parsec resolution which, via a headcount of individual SNRs, gives an
independent handle on the SFR \citep{fen08,fen10}.  \citet{mur06,mur08}
provide spatially resolved information at $\sim 1$\,kpc scale, using
Westerbork Synthesis Radio Telescope (WSRT) 20\,cm observations
\citep{bra07} combined with {\em Spitzer} 70\,$\mu$m maps. They find
that the RC emission mimics a smoothed version of the FIR map, a
consequence of CR electrons diffusing over their $\sim 10^8$\,yr
lifetime across typically $\sim 1$\,kpc. They also find that the slope
of the RC--FIR relation, $q_{70}$, varies with radius and with surface
brightness. Fig.~\ref{figure:RC-resolved} shows the stunning
similarity between 22\,cm radio continuum emission from \cite{bra07}
and a map tracing the current SFR which is based on a linear
combination of {\em GALEX} FUV and {\em Spitzer} $24 \,\mu$m emission
\cite[based on][]{hees14}. These authors, based on a sample of 18
nearby spirals, conclude that the RC--SFR relation as posited by
\citet{con92} holds for their integrated luminosities. When averaging
azimuthally, the ratio of 22\,cm surface brightness to the SFR surface
density is remarkably constant as well. Diffusion of CRe flattens the
local (averaged over a 0.7 to 1.2\,kpc diameter area) RC--SFR
relation, and  this is compatible with magnetic field
amplification by a small--scale dynamo, powered by SF-driven
turbulence.

\citet{kit14} push these results to the low-mass, low SFR regime of
dwarf irregulars (the red symbols in
Fig.~\ref{figure:RC-relation}). They find that dwarf galaxies follow
the same RC--FIR slope as that defined by the larger spirals, but are
systematically fainter by a factor of about two. Assuming equipartition,
the strength of the B--field is $9\,\mu$G, which is of the same order as
that in spirals. Resolved studies in nearby spiral galaxies
such as M51 \citep{dumas11} and NGC6946 \citep{taba13} show deviations in
the RC--IR slope as a function of environment (e.g. arm vs. inter-arm,
SFR, gas density and radiation field). These studies find evidence of
clear variations of the synchrotron emission with SFR thus providing
further evidence for coupling of the gas density and magnetic field
strength.  At SFRs lower than 1\,M$_\odot$\,yr$^{-1}$, the
RC deviates increasingly from the Condon relation, which is
interpreted as being due to a deficiency of non--thermal emission as a
result of CRe escape.

SKA1--MID will transform what can be achieved. Rather than
monochromatic (or at best in--band spectral index) information,
observations of about 1\,hr per band will sample the entire radio
continuum spectrum between 1.67 and 10\,GHz (about 4\,hr total per
target), down to $\mu$Jy sensitivity. This will allow the above
mentioned relations to be put on a much more robust footing as we will
be able to separate off the thermal 
contribution, leaving just the non--thermal fraction. In addition to
extending the work quoted above by \citet{mur11}, \citet{hees14} and \citet{kit14} to
larger samples, the spatially resolved spectral index distribution
will come within reach, which should provide a lever on the propagation
and aging of CRe as they diffuse from their sites of origin.

\subsection{Decomposition of local galaxies into their constituent parts}

Sensitive and critically high angular resolution ($\leq$0\farcs5) radio images of
nearby galaxies such as will be provided by the SKA1-MID and by the complete SKA, provide one method by which observations can
directly probe SF and SF processes in a way which is independent of complex
physical emission mechanisms. Whereas lower resolution radio
observations of normal and star-forming galaxies trace the diffuse
radio emission (see Section {\ref{localvol}}), sub-arcsecond angular
resolution observations are required to systematically characterise the populations of individual compact SF products on a galaxy
by galaxy basis by resolving away the diffuse emission. This population census can hence be
used to directly infer the levels of SF.

Observed with $<$0\farcs5 angular resolution, each individual galaxy can be considered as a laboratory containing a large
sample of discrete radio sources, all at essentially the same distance, which can be studied in a systematic way.
At $\mu$Jy and sub-$\mu$Jy sensitivities and between frequencies of 1 and 7 GHz this source population,
with the exception of accretion dominated objects and in particular
AGN (see Section \ref{accretion}), will consist exclusively of sources
related to various key phases of the stellar evolutionary
sequence. This population will be a mixture of sources from the early
stages of SF, such as compact H{\sc ii} regions, through SSCs, and
stellar end-points like X-ray binaries, planetary nebulae, SNe \citep[see also][in this proceedings]{pereztorres2014} and their SNR.

By first detecting and then identifying the physical nature of these objects using a combination of radio morphologies and spectral indices, alongside extensive multi-wavelength ancillary
data, high angular resolution radio observations will provide the first detailed extinction-free census of SF
products within nearby galaxies. The majority of core-collapse SNe
evolve to form long-lived radio SNR \citep{weiler02,gendre13}, hence this statistically well-constrained census combined with information regarding the
sizes and hence canonical ages of SNR (for the nearest galaxies observed with sub-arcsecond resolution available in higher bands with SKA1--MID), can be used to
directly infer levels of SF in individual galaxies
\cite[e.g.][]{ped01,fen08,fen10}. This will provide a further
obscuration-independent SFR tracer and, because it detects
massive stars which form CCSNe, will preferentially probe the top part
of the IMF. When combined with other SFR measures such as IR/UV or
global RC free-free and synchrotron emission
(e.g. Section\,\ref{localvol}) this will provide constraints on the universality of the
IMF as a function of galaxy type, evolution and environment within the local volume.

Importantly, such a local galaxy radio survey will also identify populations of sources that
trace earlier stages in stellar evolution, such as H{\sc ii} regions and SSCs,  placing
useful constraints on the levels of SF at various phases in the evolution of individual galaxies.
When compared with other wavelength tracers, which probe different ranges of SF age and different
spatial regions, these radio diagnostics will provide significant new insights.

Whilst this will be achievable on a galaxy-by-galaxy basis, the power
and importance of an SKA survey of nearby galaxies arises from its
large size and the available complementary multi-wavelength
data-sets. By combining these direct radio tracers of SF products, with other multi-wavelength SF
proxies (e.g. IR), significant constraints will be placed on their
calibration and interpretation, with important implications across a wide range of observational
astrophysics.  Large-area SKA1 continuum surveys with both SKA-MID and
SKA-SUR with sensitivities of $\sim$4$\mu$Jy will be capable of
detecting all local galaxies thus spanning a complete range of types and levels of both historical and ongoing SF, and allowing this census of SF products to be applied over the wide range of luminosity
and environment parameter space inhabited by galaxies. Such a survey
will provide the radio benchmark for studies of local galaxies with
application to all observational astronomers.

\subsection{Towards higher redshift}

Intermediate redshift (0.1$<$z$<$0.3) and some local Luminous Infrared
Galaxies (LIRGs) and Ultra-Luminous Infrared Galaxies (ULIRGs)  are
thought to be nearby versions of high-z star-forming galaxies
\citep[see also][in this proceedings]{mancuso2014}. 
Since a major science goal for the SKA and its pathfinders is the study of SF across cosmic 
time, it is crucial to i) have a detailed and accurate knowledge of local star-forming galaxies, and 
ii) test the radio-infrared (radio-IR) relation. Since radio emission is a dust-unbiased SF tracer,  
an accurate calibration of the radio-IR relation will be needed to determine the SFRs 
at high-z. Indeed, at $z\geq$1, 1\,arcsec corresponds to 8 kpc, so that disentangling AGN from star-forming 
activity is challenging, unless angular resolutions better than
$\sim$0\farcs1 are provided, so that one starts to separate a putative
AGN from a compact starburst at essentially any redshift. This
capability will be
provided by the full SKA.

A large fraction of massive SF at both low- and
high-$z$ has taken place in (U)LIRGs. Their implied high
SFRs are expected to result in CCSN rates a
couple of orders of magnitude higher than in normal
galaxies. Therefore, a powerful tracer for starburst activity in
(U)LIRGs is the detection of CCSNe, since the SFR is directly related
to the CCSN rate.  However, most SNe occurring in ULIRGs are optically
obscured by large amounts of dust in the nuclear starburst
environment, and have therefore remained undiscovered by (optical) SN
searches.  Fortunately, it is possible to discover these CCSNe through
high-resolution radio observations, as radio emission is free from
extinction effects.  Furthermore, CCSNe are expected (unlike thermonuclear SNe) to become strong radio emitters when the SN ejecta
interact with the CSM that was ejected by the
progenitor star before its explosion as a supernova.  Therefore, if
(U)LIRGs are starburst-dominated, bright radio SNe are expected to
occur. Given their compactness and characteristic radio behavior of
radio SNe they can be pinpointed with high-resolution, high-sensitivity radio
observations (e.g., SN 2000ft in NGC 7469 \citealt{colina01,alberdi06,pereztorres09b}; SN 2004ip in
IRAS 18293-3413, \citealt{pereztorres07}; SN 2008cs in IRAS 17138-1017,
\citealt{pereztorres08a,kankare08}; supernovae in Arp 299 \citealt{neff04},
Arp 220 \citealt{smith98,lonsdale06,parra07}, Mrk 273
\citealt{bondi05}). However, since (U)LIRGs are likely to have an AGN
contribution, high-sensitivity,
high-resolution radio observations are required to disentangle the nuclear and
stellar (mainly from young SNe) contributions to the radio emission,
thus probing the mechanisms responsible for the heating of the dust in
their (circum-)nuclear regions.

In view of the importance of (U)LIRGs in tracing the SF history across cosmic time, 
a targeted survey of local (U)LIRGs using SKA1--MID, up to a distance of 100 Mpc, will be 
essential. The sub-arcsecond angular resolution
in band 2/3 will be well-matched to current (J)VLA-A images  at higher
frequencies, permitting the thermal and non-thermal 
contribution to be disentangled in the very centres of galaxies. Considering 
the  continuum sensitivity provided by SKA1--MID, as little as 2\,minutes
per source would be sufficient to produce 1.7\,GHz images of a similar depth to those currently provided by the JVLA at 8.4\,GHz in 1\,hour. A census of all local (U)LIRGs could be obtained in just a few hours with SKA1--MID. 
Even if the specifications deviate by as much as 30\% in terms of
sensitivity, this science will 
not be severely affected. However, baseline lengths of at least 200 km
are required, to provide the necessary 
angular resolution. An SKA-MID survey providing an angular resolution
of 0\farcs5 corresponds to a physically interesting resolving linear
scale of $\sim$250\,pc at 100\,Mpc. 

Similarly, SKA1--MID will be a game--changer when it comes to providing
a benchmark study for relating the CCSN rate to the SFR in both
star-forming and normal spirals, and down to dwarf irregulars. With
sensitivities of over an order of magnitude better than current
instruments, and sub-arcsecond resolutions providing linear resolution
scales of a few tens of pc within nearby galaxies, SKA1--MID will be
well matched to spatially separate CCSNe from their surrounding
diffuse emission, thus enabling a complete census.

With the 20 fold increase in angular resolution expected
with the full SKA 
(resolution of about 10\,mas at 1.67\,GHz \citep[see also][]{paragi2014}, it will be possible to
locate individual core-collapse supernovae (or supernova remnants)
within the nuclear regions of all local starburst galaxies, similar
to detailed studies in e.g., Arp 220 and 
Arp 299 (see Fig. \ref{arp299a}, taken from \citealt{pereztorres09a,pereztorres10}), but with the potential of unveiling the much
more numerous, fainter radio supernovae and supernova remnants. In
turn, this will allow us to test scenarios of SN/CSM-ISM interaction, including estimates of the energy budgets in particles and magnetic fields, and determine the SNR luminosity vs. size relation for essentially all local (U)LIRGs. 
In addition, we will be able to extend the study described above to essentially all redshifts, 
as the 10\,mas beam at 1.67\,GHz will yield spatial resolutions of 80\,pc, or better, at all redshifts.

\begin{figure}[htbp]
\center
\includegraphics[scale=0.3]{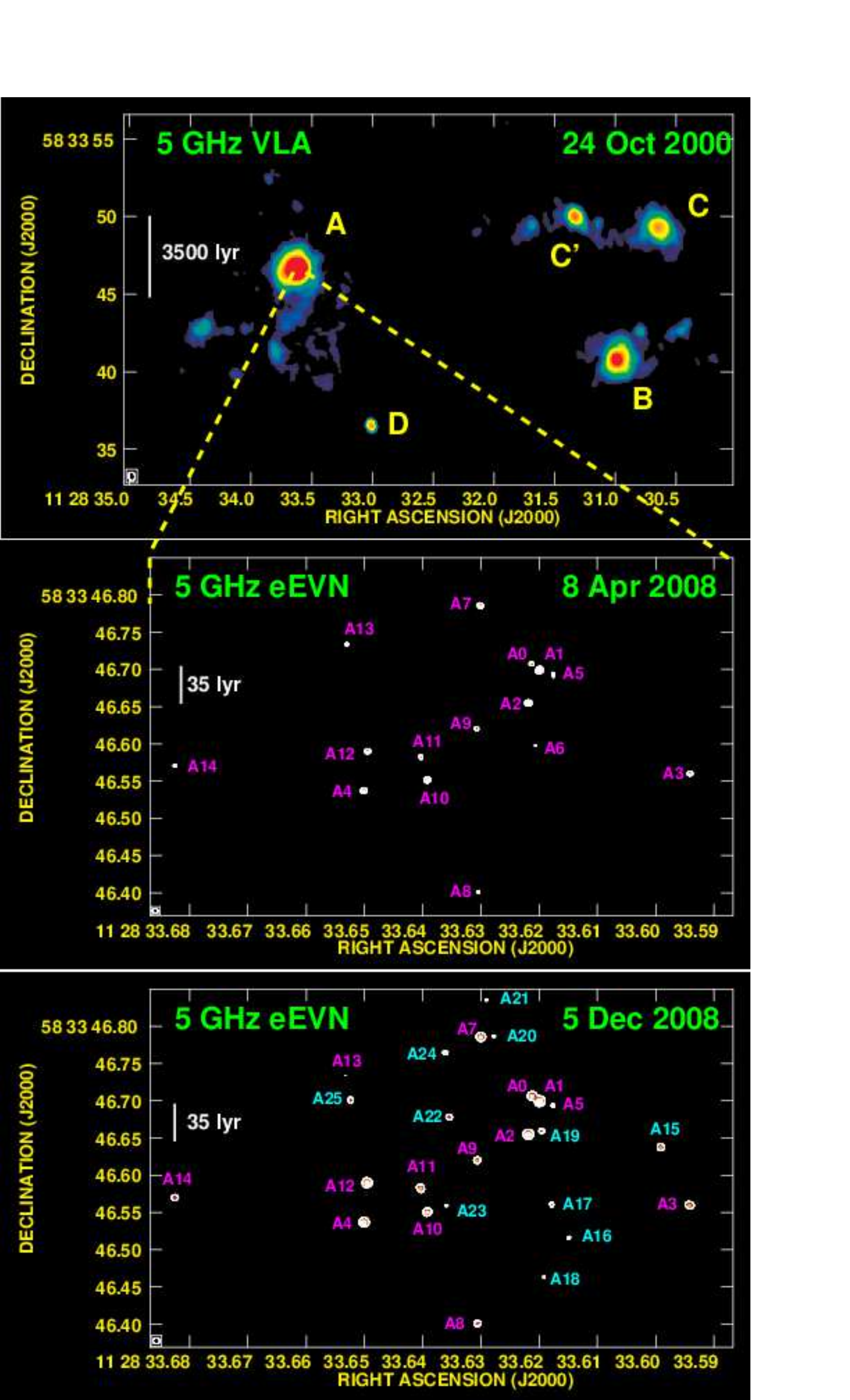}
\includegraphics[scale=0.274]{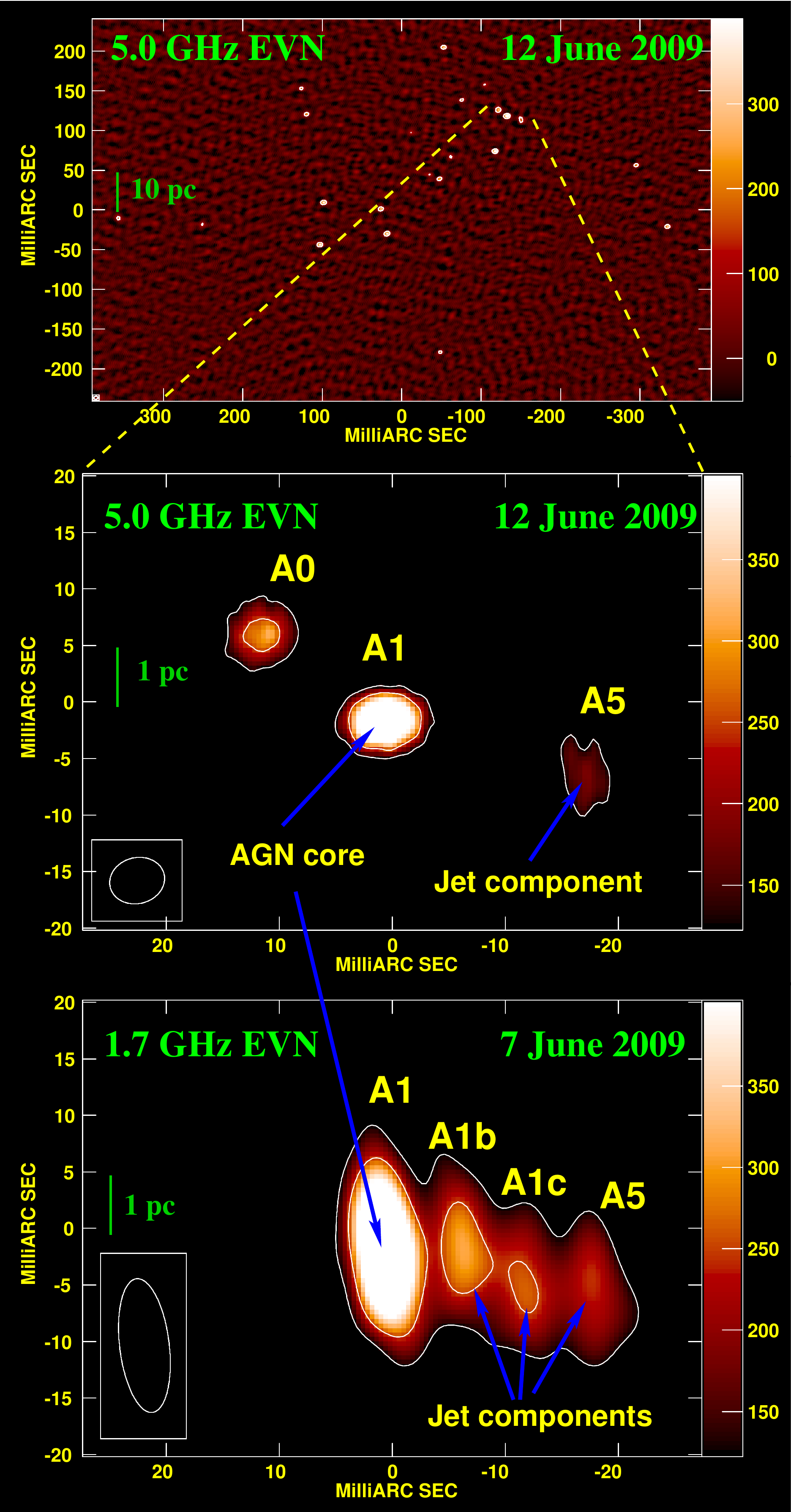} 
\caption{ \label{arp299a}
\small {\footnotesize{\it
  Top left:} 5 GHz VLA archival observations of Arp 299 on 24 October 2000,
displaying the five brightest knots of radio emission in this merging
galaxy. 
{\it  Middle and bottom left:} Contour maps drawn at five times the r.m.s. noise of our 5 GHz eEVN observations of the central 500 light
years of the luminous infrared galaxy Arp 299-A on 8 April 2008 and 5 December 2008, revealing a large population of bright, compact, non-thermal emitting sources. 
To guide the reader's eye, we show in cyan the components
 detected only at the 5 December 2008 epoch.
 {\it Top right:} 5.0 GHz full EVN image of the central 150 parsec
  region of the luminous infrared galaxy Arp 299-A, 
displaying a large number of bright, compact, non-thermal
  emitting sources, mostly identified with young RSNe and
 SNRs. 
 {\it Middle and bottom right:} Blow-ups of the inner
 8 parsec of the nuclear region of Arp 299-A, as imaged with the full
 EVN at 1.7 and 5.0~GHz.  
 The core-jet morphology, spectral
 index and luminosity of the A1--A5 region clearly revealed the location 
  of the long-sought AGN in Arp 299A. 
 }}
\end{figure}

\section{Accretion sources in the nearest galaxies}
\label{accretion}
%RJB

The sensitivity and frequency coverage of SKA1 will enable a
complete census of accretion powered sources in the local
Universe. Such sources encompass multiple orders of magnitude in mass
and luminosity, from supermassive black-holes (SMBH) and low-luminosity AGN, through Sgr\,A$*$--like luminosities, to intermediate-mass
black-holes and down to stellar mass black-hole systems such as
microquasars. The sensitivity of the SKA during a deep pointed survey
will show accretion-dominated objects in a variety of
environments within an individual galaxy, not constrained just to
the nucleus. 

\subsection{Accretion onto supermassive Black-holes}

Accretion onto SMBHs is one of the most significant energy sources in
the universe, with the potential to clear star-forming gas from
galactic bulges and even to regulate the growth of entire galaxies in
galaxy clusters \cite[e.g.][]{dimat05}. The mechanism for this feedback
is mechanical, through jets and outflows powered by
accretion. However, despite the importance of SMBH activity in
regulating galaxy formation, comparatively little is known about SMBH
activity towards low radiative luminosities. This is a significant gap
in our understanding of feedback and the role of SMBHs in galaxy growth
and evolution, since it is now known that mechanical jet power can be
energetically more significant than supernova feedback, even at low AGN
luminosities \citep{nagar05, koerding08}. The major
difficulty in studying low luminosity AGN is precisely their low
radiative output with respect to their surrounding host galaxy,
especially since many LLAGN are embedded in nuclear SF regions 
\citep[see Fig.\,\ref{arp299a}; and work of][and references
  therein]{ho97}. 
For these LLAGN, large amounts of mechanical
energy are shed by massive stars into their surrounding medium which
significantly increases its temperature. Thus those massive stars would hinder
the accretion of material to the central black-hole, resulting in a
less powerful AGN.  
 The problem is compounded because at low-luminosities AGN
become radiatively inefficient, and lose the strong optical and X-ray
signatures commonly used to identify AGN activity. However, just
because an AGN is radiatively inefficient does not make it
mechanically inefficient, in fact at low luminosities the total power
is almost certainly mechanically-dominated via the jet.

Radio imaging is also essential to unambiguously identify activity due to
SMBH accretion. For example, even with more sensitive optical
spectroscopy, it will be difficult to distinguish the weak optical emission from faint AGN against the contribution from nuclear SF. In contrast, a separation of AGN and SF components can be made with the
moderate resolution ($\sim$0\farcs5) obtained by SKA1--MID initially in
the most local systems, coupled with the leverage of multiple frequency bands to distinguish flat-spectrum
AGN cores. Finally, using radio data circumvents the problem of where to draw
the dividing line between a galaxy being ``active'' or ``normal''. This is
because one can not only estimate the jet power from core radio luminosities, but also measure the absolute mass accretion rate, which
appears to be well-correlated with radio luminosity for the lowest
luminosity objects, and where beaming effects do not play an important
role, scaling as L$_{\rm R}$ $\propto$ M$_{\odot}^{1.4}$ \citep{koerding06}. Thus we can
classify all galaxies according to their SMBH accretion power, tying
the lowest accretion ``quiescent'' SMBH to the LINERs and classical
Seyferts. 

\subsection{Accretion from intermediate to stellar mass black-holes}

Deep radio surveys of the local Universe with the SKA will prove
critically important to our further understanding of high-energy accretion
processes across a wide mass-function of compact objects. Bright
off-nuclear X-ray point sources with luminosities in excess of
$10^{39}$ erg~s$^{-1}$, known as Ultra-Luminous X-ray sources (ULXs),
are among the most energetic mass-accretion processes in the local
Universe \citep{roberts07}. Due to their defining characteristics, ULXs
could be explained by stellar-mass black-holes with global accretion rates at
or in excess of their Eddington limit or represent a population of
intermediate mass black-holes (IMBH) with masses between
$10^{2-3}$M$_{\sun}$ \citep[e.g.,][]{king09}. Some evidence for the
latter has arisen through the discovery of radio emitting `bubble
nebulae', 100s of parsecs in diameter surrounding the X-ray sources
\citep[e.g.,][]{pakull02}, and transient radio outbursts from a
`hyper-Luminous X-ray' source at a distance of $\sim100$~Mpc \citep{webb12}. These may be photo-ionised by the central X-ray source
\citep{pakull02,kaaret04}, shock-heated through the
interaction of outflows with the ambient medium \citep{pakull02,pakull03} or both. As the bolometric radiative efficiency of these
sources provides an unreliable measure of the black-hole mass, it is by
studying these surrounding radio nebulae and jets that we could
estimate the true kinetic power of these compact objects and gain
insight to their mass-function \cite[see more extensive chapter on
  this area by][]{wolter2014}.

IMBHs and stellar-mass black-hole systems (e.g. microquasars) also provide a hitherto
sparsely investigated population of transient or variable faint
compact radio sources in galaxies and are expected in larger numbers
in high SFR systems than are found in our own Galaxy \citep[see][for more information on radio transients and the SKA]{fender14,corbel14}. To date few dedicated radio searches have been performed of nearby
star-forming galaxies and only a handful of candidate sources have
been detected serendipitously \citep[e.g., potenial microquasars in M82,
  Arp220 and Arp299;][respectively]{mux10,jos11,bat12,bondi12}. Whilst
the current radio instruments, such as JVLA and e-MERLIN, are only just starting to explore the brightest and nearest of this
population. Via large-area and pointed SKA survey observations with
sub-arcsecond angular resolution the entire population of accreting
objects in galaxies across all mass scales and environments will be charactised.
Searches for radio emission from star clusters and dwarf satellite
galaxies around local galaxies will provide constraints on what
fraction of each class of object shows evidence for containing black-holes of
$\sim$1/1000 of their total masses, in the same way that giant
galaxies seem to do nearly universally. Several visits to the
fields my be required to clarify the nature of the sources.

\section{Tracing the fuel for, and influence of SF and accretion: Molecular
  Gas tracers of kinematics and properties}
\label{gas}
\label{masers}
\subsection{OH masers and Circumnuclear starbursts}

Masers are the radio analogues of lasers, and occur when there is an
excess of molecules in a higher energy state -- a non-thermal or
inverted population distribution.  This gives a negative optical depth
so that ambient or background radiation is amplified. Intrinsically
compact both spatially and
spectrally  (extragalactic masers are detectable down to $\mu$as
scales in $<$km\,s$^{-1}$ channels), masers provide the best
directly-mappable tracers of high-resolution kinematics.  They
demonstrate the presence of more compact molecular regions than any
other cm-wave tracer and increasingly
sophisticated models (e.g.$\!$ \citealt{Gray12}) place tight limits on the
density-temperature parameter space required to pump the maser inversion.
The most intense Milky Way masers are associated with individual star
forming regions, and similar phenomena have been detected from 
Local Group galaxies, \citep[water, methanol and hydroxl;][respectively]{Brunthaler06,Sjouwerman10,Argo13}.  So called Mega-
or kilo-masers can be more than a thousand time brighter and hence can
be detected from within more distant
galaxies \citep[see][for a recent review]{Tarchi12}.  Some of the
potential science applications of
the SKA for investigating nearby and Galactic masers is covered by
\citet{etoka2014} and \citet{robishaw2014} in this proceedings.

OH masers, rest frequencies 1.6-1.7 GHz, have been found in over 100
(U)LIRGs out to $z$=0.27 \citep{Darling02,Willett12}.  There
is a strong relationship $L_{\mathrm{OH}} \propto
(L_{\mathrm{IR}})^{\alpha}$, explained by the r\^{o}le of radiation from
dust in the maser pumping \citep{Lockett08}. The corresponding OH IR
lines at 35 and 53$\mu$m have recently been detected in both
absorption (53$\mu$m) and emission by \emph{Herschel}
\citep{Gonzalez-Alfonso14} providing important constraints on pumping
schemes. Values of $0.5<\alpha<2$ have been reported,
however these are sensitive to the angular resolution of the observations and the
orientation of the emission region. Dust temperatures $45<T<200$\,K
(optimally 80--140 K) and a number density $10^{9}<n<10^{12}$\,m$^{-3}$
(\citealt{Darling07,Lockett08}) are also required.

Images of extra-galactic OH masers have only been published for around a dozen galaxies
further away than the Magellanic Clouds, out to $z\sim0.045$. In
approximate order of increasing distance, these are: M82
\citep{Argo10,Argo13}, NGC\,1068 \citep{Gallimore96}, III\,Zw35
\citep{Parra05}, Arp\,220 \citep{Rovilos03}, Arp\,299
\citep{Polatidis01}, II\,Zw096 \citep{Migenes11}, Mrk\,273
\citep{Yates00}, Mrk\,213 \citep{Richards05}, IRAS\,17208-0014
\citep{Momjian06}, IRAS\,12032+1707, IRAS\,1407+0525 \citep{pihlstrom05} and IRAS\,10173+0828 \citep{Yu05}. All these galaxies
show evidence for interactions or mergers and some masers are found at
two or more sites hundreds of pc apart \citep[e.g., in Arp220;][]{Rovilos03}.

In most of the more distant objects, the brightest masers trace a
rotating disc, typically a (few) hundred mas (a few hundred pc) in
diameter, suggesting enclosed masses between $10^6-10^9$\,M$_{\odot}$,
possibly evidence for massive black-holes, but dense nuclear starburst
regions, or both \citep{Richards05}, cannot be ruled out without
higher-resolution data \citep{Klockner03}.  These Mega-masers are
generally inside the extent of discs traced by H{\sc i}.  VLBI observations,
and modelling of lower resolution data, show that the OH masers mostly
emanate from clouds within a factor of four of 1pc in diameter
\citep{Lonsdale03,Richards05}.  The discs may also be
warped or interact with jets, requiring sensitive high-resolution
imaging for good models \citep[e.g.,][]{Klockner03}.

The OH ground state main lines, rest frequencies 1667.359 and
1665.402-MHz are brightest, with intensity ratio $\ge1.8$ (the value
expected for unsaturated emission) and (in accordance with
predictions), the highest brightness temperature masers have higher
ratios \citep{Lonsdale98}.  The OH satellite lines at 1612- and
1720-MHz are fainter and masers have only been detected in 6 objects
\citep{McBride13}. There are cases where 1720-MHz absorption
accompanies 1612-MHz emission, requiring a different pumping mechanism
even if mainline masers are also seen.  A number of OH Mega-masers have
line wings at up to 800\,km\,s$^{-1}$ from the systemic velocity
\citep{Baan89}, but neither these nor the satellite lines have ever
been imaged.

\subsection{Other cm-wave extragalactic masers}

Bright 22-GHz extragalactic masers are best known thanks to their
location in sub-pc tori around supermassive black-holes.  Both Doppler
and (with VLBI) proper motions were measured for NGC\,4258
\citep{1999Natur.400..539H,green2014,paragi2014}.  This gives the most precisely-known
estimates of the enclosed black-hole mass and the distance, providing
the disc kinematics are modelled accurately
\citep{2013ApJ...775...13H}.  The Hubble constant could thus be
measured very accurately from a larger sample of such galaxies \citep[e.g., the
Mega-maser cosmology project][]{Kuo13}. VLBI with the SKA will push this to far greater distances.
%, see\citet{green2014}
A water maser at
$z$=2.6 has already been detected at 6.1\,GHz \citep{Impellizzeri08} in
a lensed galaxy.  H$_2$O masers are also found in interactions between
jets and the ISM, \citep[e.g. ][]{Kondratko05}.  Such galaxies do not
usually have high SFRs and only a few nearby merger
systems, e.g. NGC\,1068, also host OH masers.

Excited OH lines occur in the rest frequency range 4-6.7\,GHz;
extragalactic thermal emission is known \citep{Impellizzeri06},
but as yet these lines have not been observed as masers.  Formaldehyde masers (rest frequency 4.8\,GHz)
are known from a handful of ULIRGs, all with OH Mega-masers \citep{Araya04}.
The inversion has the interesting property that maser amplification
occurs at $n\ge10^{11.6}$\,m$^{-3}$ and the excitation temperature
exceeds the CMB temperature; at lower densities supercooling and
strong absorption occurs \citep{Mangum13}. Molecules, such as OH and CH, with multiple
masing transitions can be not only good thermometers, but also
can be used as probes of any cosmological variation in fundamental constants.

\subsection{SKA phase 1 and SKA observations of Mega-maser molecules}

There are several complementary modes whereby SKA1--MID can locate and
measure Mega-maser properties. Spectral surveys would be very valuable
in finding Mega-masers associated with galaxies out to the peak of the
merger rate around redshift 1--2, helping to provide an estimate of the galaxy
merger rate \citep{aharonian13}, and in identifying candidates for
imaging.  Searches are optimised by providing at least 3 samples
across the line peak, i.e. a few tens km\,s$^{-1}$ channels,
with angular resolution $\le1$ arcsec, not greatly exceeding the size
of the emission area.  Nonetheless, piggyback observations at coarser
resolution would be worthwhile. Selection criteria are described based
on IR properties \citep{Darling02} including deep silicate 9.8$\mu$m
absorption \citep{Willett11}. The 18 sources detected at $0.1<z<0.27$
have OH peaks of 1.8--40\,mJy and line widths (FWHM) of 50--600\,km\,s$^{-1}$, with some of the most distant being among the brightest.
This suggests that a few percent of similarly selected candidates
would have a flux density $\sim$0.6 mJy at $z=1$ and the brightest
might reach $\sim$0.2 mJy at $z=2$.  Assuming that a sensitivity of
0.13\,mJy can be reached in 20\,km\,s$^{-1}$ channels in $\le1$ hr
(depending on Dec.) at 1 arcsec resolution, this would give a
5$\sigma$ detection for 0.6 mJy masers in 1-2 hr, at modest redshifts
around 1.6 GHz. At 555 MHz, $z\sim2$, the lower sensitivity requires
$\sim24$ hr to detect 0.2 mJy lines. \cite{McKean09} describe a
strategy using APERTIF or ASKAP for a blind search which would detect
$\sim1$ mJy masers ($5\sigma$) at the highest accessible redshift of
$z=1.39$, but in several hours (per tuning) and 150\,km\,s$^{-1}$ resolution.
 
The known satellite lines are 1--10\% of the main line peaks in
Mega-maser galaxies, but are often brighter in Milky-Way-like
star-forming regions.  \citet{McBride13} detected 5 out of 77
sources at $z<0.05$, at a few mJy.  The 1-hr sensitivity of SKA will
go at least twice as deep. Detection of multiple maser lines places
tighter constraints on the parameter space of
temperature-density-velocity coherence.  Such a survey may also produce a large
enough sample of satellite maser lines to help identify the pumping
mechanism responsible for conjugate behavior, especially for any
bright enough to image.  

The enclosed mass density derived from resolved maser PV diagrams can
be compared with predictions from thermal tracers of high-density,
warm gas, for example HCN, which has excitational properties correlated
with OH Mega-masers \citep[][and see Section \ref{otherlines}]{Darling07,Kandalyan10} and with HC$_{3}$N \citep{Lindberg11}.  The position of
maser components can be measured with a precision (for reasonable uv
coverage) of 0.5$\times$(beamsize/signal-to-noise ratio), so, at $250$\,mas
resolution, a signal-to-noise of 5 across a few hundred km\,s$^{-1}$, sampled by 5
or more channels, would resolve a $\sim150$\,pc radius edge-on disc at
$z\sim0.2$. The 2-hr sensitivity at this resolution in 30 km\,s$^{-1}$
channels is $\sim0.5$\,mJy, allowing emission above 2.5 mJy to be
resolved sufficiently.  This would typically correspond to a 10 mJy
spectral peak.

Selected objects should be imaged for 12--24 hr, to allow higher
sensitivity or higher resolution.  This will allow detection of
multiple components per channel from discs tilted enough to be
spatially resolved, and define the angle of inclination.  The 1665-MHz
line is typically fainter and lacks the compact hot-spots seen in
the 1667-MHz line; imaging this avoids blending ambiguities as well as
providing material for maser modelling.  Higher sensitivity will provide
the first imaging of the satellite lines, providing tighter constrains
on physical conditions.  It will also reveal the origin of the
high-velocity mainline emission, which could be fast-rotating regions
in the inner disc, or material entrained by jets \citep{Klockner03},
since at least some of the host galaxies contain weak AGN, although
the fraction has been variously estimated at 45\% \citep{Baan98}
or 10--25\% \citep{Willett11}.  Another possibility is a molecular
outflow as suggested by the velocity dispersion within the III\,Zw35
disc \citep{Parra05} since the superwind from NGC\,253 has recently
been shown by ALMA to be laden with molecules as well as the
better-known light, ionised component \citep{Walter13}.

Searches and rapid imaging of extragalactic masers requires a channel
width of a few tens of km\,s$^{-1}$ at most, since maser lines tend to be
narrow and diluted at too-coarse a resolution.  Spectra can be
decomposed into a mixture of broad features and few km\,s$^{-1}$
details, and a resolution of $\sim1$ km\,s$^{-1}$ is needed to measure
Zeeman splitting \citep{robishaw2014}, or even finer for the measurement of fundamental constants.  Similarly, comparison of single dish,
WSRT, MERLIN and EVN spectra \citep{Klockner04} shows that the
brightest peaks are often almost as bright at $\sim30$\,mas resolution 
as at a resolution of a few arcsec or more, but the weaker emission is
partly or completely resolved out on scales of a few hundred mas or
less. The SKA will provide the first opportunity to sample emission on
MERLIN to WSRT scales simultaneously (and eventually on VLBI scales).
Flexible data reduction is needed, with weighting-up of short
baselines for maximum sensitivity at high spectral resolution, and
channel averaging to allow high spatial resolution making full use of
long baselines.

Finally, another motive for characterising OH masers is to avoid
contamination of H{\sc i} surveys; the maser lines can be distinguished at
higher spectral resolution or by anomalous redshift and excess
brightness with respect to other lines and continuum/optical
properties \citep{Haynes11}.

\subsection{Summary of Mega-maser requirements}
The basic requirement is the frequency range.  Ground-state OH masers
at 0$<z<2$ need frequencies from about 1721 down to 530 MHz.
Searches are optimized at a velocity resolution of ten or a few tens
of km\,s$^{-1}$. Follow-up of brighter objects at resolution down to
$\le1$ km\,s$^{-1}$ in full polarization is desirable, with a coverage
of a few thousand km\,s$^{-1}$ and angular resolution as fine as
possible (ideally 100 mas or less).  However, it will not be possible
to achieve many detections simultaneously at full spectral and spatial
resolution. To achieve this the use of different weighting may be
required to produce the highest spectral or spatial resolution
images.

The (to date) rarer extragalactic masers such as CH, formaldehyde,
methanol and excited OH require frequencies up to just above 6 GHz
(potentially higher, since Galactic OH and methanol are known up to 13\,GHz). In fact, any expansion of frequency range will allow detection
of masers at some redshift, notably for the 22-GHz water masers.  In
the more advanced stages of the SKA, VLBI at up to 22 GHz is essential
for the Megamaser Cosmology Project.

\subsection{Molecular line and other tracers}
\label{otherlines}
There is now mounting evidence that cold, neutral/molecular outflows are
extremely important carriers of momentum and mass within galaxies. We are beginning to see that the
properties and mass balance atomic/molecular/ionised change along individual outflows as well
as among outflows. In this area H{\sc i} absorption studies are
absolutely fundamental, however observations of OH maser emission in
dense outflows in ULIRGs also have a critical role and radio
recombination lines (RRLs) also potential can play an important part
in nearby sources \citep[see
section\,\ref{masers} and][for H{\sc i} absorption and RRL opportunities with
  the SKA]{morganti2014,oonk2014}.
Very recently it has also become clear that AGN driven outflows may also have extremely interesting
molecular properties (linked to their evolution) that can be studied
in exquisite detail with the SKA. The ISM properties of galaxies throughout the Universe can be
investigated via H{\sc i} and radio recombination studies which probe the atomic and ionized medium
and are a key complement to the molecular cloud
structure/mass/dynamics probed by mm-telescopes such as ALMA. This
area is of fundamental importance to the understanding of cloud and
SF processes in galaxies both near and far.

In this area the sensitivity and frequency coverage of the SKA, even
in phase-1, provides a unique opportunity to study these processes via
centimetric molecular lines that have hitherto been ill explored
\cite[preliminary JVLA and Arceibo studies have been made on selected
  sources e.g.,][]{ric11,salter08}. A
number of molecular tracers, which include several prebiotic molecules, are available within the proposed SKA-1
bands (see Table\,\ref{mol-lines}) and will be observable as kinematic
and ISM tracers within local galaxies.

For example the SKA's capabilities provide the possibility of studying the properties of the extreme nuclear regions of Compact
Obscured Nuclei (CONs) - and dust obscured galaxies (DOGs). Here, activity may be buried
behind N$_{\rm H}> 1\times10^{25}$\,cm$^{-2}$ of gas which means that these Compton thick, high pressure regions are the
realm of molecules (not even X-rays will penetrate to help fingerprint the activity). Molecular
emission can probe behind the optically thick dust veil to reveal temperature gradients and field
strengths at SKA spatial resolution. Thus these studies will trace the impact and structure of the
deeply buried activity in a unique way. The properties of these
galactic nuclei may only be accessible with
molecules.

\begin{table*}
\caption{Selection of molecular line tracers between 1-10\,GHz and
  within the bands of SKA-1. Lines of the OH radical and various other species are also of key importance
  but for brevity not included below.}
\scalebox{1}{
\begin{tabular}{lccc}
\hline
Line    & Rest freq (MHz) & SKA-1 MID band & SKA-1-SUR band \cr
\hline
HCN $v_2=1, \Delta J=0,J=2$ & 1346.765& 2 &  2 \cr
HCN $v_2=1, \Delta J=0,J=4$ & 4488.4718& 4 &  - \cr
H$_2$CO & 4829 & 4/5 & - \cr
N$_2$H+ & 5009.8278& 4/5 & - \cr
H$_2$CNH 1$_{10} - 1_{11}$,$\Delta F=0,\pm1$ & 5289.813 & 5 & - \cr
HCO+ & 6350.908& 5 & - \cr
HNC & 6484.497& 5 & - \cr
CH$_3$OH 5$_1-6_0 A^+$& 6668.5192 & 5 & - \cr
HCN $v_2=1, \Delta J=0,J=5$ & 6731.9098& 5 &  - \cr
HCO+ & 8890.452& 5 & - \cr
HCN $v_2=1, \Delta J=0,J=6$&9423.3338 & 5 & -\cr
HNC & 9724.644& 5 & - \cr

\hline
\end{tabular}
}
\label{mol-lines}
\end{table*}

Why SKA and not ALMA? Many of these line species are accessible at mm and
submm wavelengths (and studies are underway with ALMA). However, some of the heavier molecules have their primary transitions
in the SKA bands, plus critically at submm wavelengths the line confusion and line blending is so severe that
using these lines sometimes becomes extremely difficult. Pilot studies
on nearby galaxies are underway on the JVLA at 6, 9 and 20\,GHz,
however these will remain sensitivity limited with current
facilities. The sensitivity and frequency range covered by the SKA
make this it an ideal instrument to open up this new field.
Furthermore initial modelling work predicts that several of these molecules will have maser transitions at the
very lowest frequencies (e.g. 0.3\,GHz) rendering them as potentially extremely important probes of circumnuclear
extreme regions around AGNs.

\section{Conclusions}

The major endeavor that is the SKA will create a general user facility
which will produce transformational science across a wide range of
astrophysics for many decades to come. For studies of the galaxy
population within the local Universe these broad, general user
capabilities, will enable an extremely wide range of science covering many areas of astrophysics, and will form the
bridge between the detailed studies of objects in our own Galaxy and the
distant high-redshift Universe. Deep and moderate-resolution (few arcsec)
continuum and spectral line surveys of large numbers of nearby galaxies will
be provided by projected SKA1 ``all-sky'' surveys at frequencies of
$\sim$1-2\,GHz. Such large area surveys will essentially provide a
full radio atlas of the local Universe and allow detailed studies of
the non-thermal radio component of galaxies. However, multiple, pointed observations
covering a wide range of frequency bands,
in particular higher band SKA1--MID (band 5), will be required to
characterise the non-thermal components of local galaxies. Even with
modest capability reductions, SKA1 will make significant
advances in this field. This will, for the first time, allow the determination of SF within
galaxies covering the full range of type and environments which will
be critical to our understanding of galaxy evolution and
SF through cosmic time. 

Via higher frequency bands in SKA1 and later baseline extensions during
the full SKA high angular resolution ($>0\farcs5$ to mas) continuum and
spectral line (maser and absorption studies) of nearby galaxies
will allow the physical composition of these galaxies to be
decomposed. Even with the modest resolution of SKA1, this will
result in the detection and study of many thousands of individual
SF and accretion-related components such as SNe, SNR,
H{\sc ii} and their like. Via such observations the SKA will not only be
able to study the physics of SF and extreme physics in
accretion dominated sources on an individual basis in nearby galaxies,
but also allow statistical properties of these sources to be
investigated, how they interact with the ISM and how they affect
galaxy evolution.

\bibliographystyle{apalike}

\end{document}